\documentclass[a4paper,12pt]{article}

\usepackage{graphicx}
\usepackage{amsmath,amssymb}
\usepackage{a4wide}
\usepackage{cite}

\begin{document}

\begin{center}

   {\large \bf Ab initio study of element segregation and oxygen adsorption \\
                       on PtPd and CoCr binary alloy surfaces}

   \vspace{1cm}
    Arezoo Dianat,$^{1}$
    Janina Zimmermann,$^{2}$
    Nicola Seriani,$^{3}$ \\
    Manfred Bobeth,$^{1}$
    Wolfgang Pompe,$^{1}$
    and Lucio Colombi Ciacchi$^{2,4}$

   \vspace{0.25cm}
   {\it
       $^1$Institut f\"ur Werkstoffwissenschaft, Technische Universit\"at Dresden, \\
           Hallwachsstrasse 3, 01069 Dresden, Germany \\
       $^2$Fraunhofer Institut f\"ur Werkstoffmechanik \\
           W\"ohlerstrasse 11, 79108 Freiburg, Germany \\
       $^3$Institut f\"ur Materialphysik, Universit\"at Wien, \\
           Sensengasse 8, 1090 Wien, Austria \\
       $^4$Institut f\"ur Zuverl\"assigkeit von Bauteilen und Systemen, \\
           University of Karlsruhe, Kaiserstr. 12, 76131 Karlsruhe, Germany. }

\end{center}

\vspace{0.5cm}
\begin{center}
  \today 
\end{center}

\newpage
  \begin{center}
     {\bf Abstract}
  \end{center}
The segregation behavior of the bimetallic alloys PtPd and CoCr 
in the case of bare surfaces and in the presence of an oxygen ad-layer  
has been studied  by means of first-principles modeling based on 
density-functional theory (DFT).
For both systems, change of the d-band filling due to charge transfer
between the alloy components, resulting in a shift of the d-band center of surface 
atoms compared to the pure components, drives the surface segregation and 
governs the chemical reactivity of the bimetals. 
In contrast to previous findings but consistent with analogous PtNi alloy systems,
enrichment of Pt atoms in the surface layer and of Pd atoms in the first subsurface 
layer has been found in Pt-rich PtPd alloy, despite the lower surface energy 
of pure Pd compared to pure Pt.  
Similarly, Co surface and Cr subsurface segregation occurs in Co-rich CoCr alloys.
However, in the presence of adsorbed oxygen, Pd and Cr occupy preferentially 
surface sites due to their lower electronegativity and thus stronger oxygen affinity
compared to Pt and Co, respectively.
In either cases, the calculated oxygen adsorption energies on the alloy surfaces are larger
than on the pure components when the more noble components are present in the 
subsurface layers.
\\
\\
\\
\\
Keywords: electronic structure calculations, bimetallic surfaces, alloying, 
palladium, platinum, chromium, cobalt, chemisorption

\newpage

\section{Introduction}

Mixing of two metal elements to form an alloy permits to tune to a certain 
degree the thermal, mechanical and electronic properties of the resulting material.
In particular, the surface reactivity of an alloy, often mediated by an oxide 
layer forming in contact with the atmosphere, may differ substantially from 
the reactivities of its separate components.
This is due to the peculiar surface electronic properties resulting from 
selective segregation processes of the alloying elements near to the surface.
For instance, bimetallic alloys may exhibit superior catalytic properties compared to 
pure elements~\cite{Lopez,Rodriguez,Gross2006,Stamenkovic_2007,Ye_2007}.
In particular, mixtures of platinum and palladium are widely applied 
for hydrogen reactions~\cite{McKee,Ruiz,Deng2,Dedanello,Moysan} and
oxidation~\cite{Skolg,Vigneron}, as well as for electrochemical oxygen
reduction reactions~\cite{Stamenkovic_2007,Ye_2007}.
Furthermore, the surface of a metal prone to corrosion may be passivated by alloying 
with elements which selectively form a stable oxide layer, as in the case of FeCr or
CoCr alloys used in biomedical applications~\cite{Dearnley_1998}.
Both in the case of noble metal alloys used in catalysis and of non-noble metal alloys 
designed to sustain the corrosive attack of an oxidizing environment,  
detailed knowledge of the surface composition and reactivity is necessary 
to rationalize and optimize the alloy performance.
In either case, the surface properties are heavily influenced by the reaction of the 
bare surfaces with oxygen.

In this work, we aim to gain basic knowledge of the details of the cooperative
interactions among the elements composing both noble and non-noble 
bimetallic alloys in proximity of their surfaces.
On one hand, the interactions of the elements in the alloy (e.~g.~resulting
in charge transfer between the atoms) may influence the surface reactivity
towards oxygen adsorption.
On the other hand, the formation of an oxygen ad-layer may in turn have an effect
on the alloy composition just beneath the surface.
To address these issues, we study the formation of an oxygen ad-layer on the 
surfaces of bimetallic PtPd and CoCr alloys by means of first-principles modeling
based on density-functional theory.

Both the PtPd and the CoCr systems exhibit negligible alloying stress due to 
the very  similar lattice constants of their components.
Namely, as obtained in our calculations, the lattice constant of Pt is only 0.025~\AA\
larger than that of Pd, and the Co-Co nearest neighbor distance in hcp-Co ($\sim$2.49~\AA) 
is very close to the Cr-Cr distance in bcc-Cr ($\sim$2.48~\AA).
Thus, for ideal PtPd and CoCr crystals, surface segregation is expected to be 
driven mainly by minimization of the surface energy (which might include the 
effect of adsorbed oxygen) as well as the mixing free energy.

In the last couple of decades, the Pt-Pd phase diagram has been studied 
experimentally~\cite{Kidron,Hayes,Darby,Bhara2,Bhara1,Noh} as well as
theoretically~\cite{Lu2,Lu1,Kolb,Turchi}.
Calorimetric~\cite{Hayes} and x-ray diffuse scattering~\cite{Kidron} 
investigations have shown that PtPd alloys exhibit negative mixing enthalpies 
over the whole composition range. 
Theoretical investigations with cluster expansion methods are consistent with this
finding and predict ordering in an L1$_0$ phase at a composition of 50~at.\%~\cite{Lu1,Kolb}.
As far as the surface composition of PtPd alloys is concerned, existing theoretical and
experimental works indicate preferential Pd segregation both for Pt-rich and Pd-rich systems, 
to an extent depending on the crystallographic surface orientation~\cite{Rousset1}.
It has to be noted that experimental measurements at room temperature
may be affected by metastable frozen-in surface configurations.
For instance,  using low-energy ion scattering~\cite{Oetelaar} segregated surface 
compositions of Pt$_{20}$Pd$_{80}$ and Pt$_{80}$Pd$_{20}$  alloys were found 
to be in equilibrium only at temperatures above 700 K.
Available theoretical results on the surface composition of PtPd nanoparticles have been obtained 
either with analytic embedded-atom methods~\cite{Deng}, empirical~\cite{Tomanek} 
or semiempirical potentials~\cite{Massen}, indicating preferential Pd segregation in all
cases.
This seems to contrast with the findings of recent combined experimental and theoretical
studies based on density-functional theory on Pt alloys containing, as is the case of
PtPd, a less electronegative element such as Co or Ni (the Pauling electronegativities
of Pt, Pd, Ni and Co are 2.28, 2.20, 1.91 and 1.88, respectively~\cite{Webelements}).
In PtCo or PtNi alloys, the less electronegative elements with respect to Pt are found 
to segregate to subsurface sites, leading to a Pt-rich surface (a ``Pt skin'') 
with highly enhanced catalytic activity with respect to pure Pt (see e.\ g.\ 
Ref.~\cite{Stamenkovic_2007} and references therein).
This raises the question, whether subtle electronic effects which can only be captured
by accurate quantum mechanical modeling may be responsible for analogous Pt surface 
segregation effects in PtPd alloys.
These effects may be related to the recently observed high catalytic activity of PtPd nanoparticles 
with a Pd concentration  between 17 and 33\,\%~\cite{Ye_2007}.

Compared with the PtPd system, less information is available on the surface composition 
and the segregation behavior of bare CoCr alloys.
It is well known that under atmospheric conditions a CrO$_{x}$ passivation layer
containing a minority of Co and Mo develops on the surfaces of technological 
CoCrMo alloys~\cite{Milosev}.
This points towards surface segregation of Cr in an oxidizing environment.
However, to our knowledge, no experimental or theoretical study has been performed
to date to investigate the segregation behavior of CoCr alloys under vacuum conditions.

In the following sections, after a brief description of the computational technique,
we report on results of total energy calculations with structural optimization of
alloy models with different surface compositions, both in the absence and in the presence 
of adsorbed oxygen.
In Section 4, the calculated adsorption energies of oxygen on the alloy surfaces are presented
and rationalized within the so-called d-band center model by Hammer and N{\o}rskov~\cite{Hammer}.
Finally, the main conclusions of the work are summarized in Section 5.

\section{Computational Details}

The calculations presented in this paper have been performed within the framework of 
the spin-polarized density-functional theory (DFT) using the generalized gradient 
approximation (GGA) to the exchange-correlation functional and periodic boundary 
conditions.
The projector augmented wave (PAW) method~\cite{Blochl,Kresse2} has been used 
to represent the interaction between valence electrons and core ions, as implemented 
in the VASP~\cite{Kresse3,Kresse1,Kresse4} and  LAUTREC~\cite{Lautrec} codes. 
Total energy differences computed with the two codes were found to
differ by no more than 20~meV per atom, provided that equivalent PAW
datasets and similar exchange-correlation (xc) functionals were used.
In particular, our results for the PtPd system have been obtained
with the VASP code using the PBE xc functional and for the CoCr system 
with the LAUTREC code using the PW91 xc functional.

Both Pt and Pd PAW datasets have been generated with 10 explicit 
valence electrons, while 9 valence electrons have been included
in the dataset for Co.
For these three elements, 3, 2 and 2 projectors have been used 
for the s, p and d angular momentum channels, respectively.
Our Cr dataset treats explicitly 8 semi-core and 6 valence
electrons (corresponding to the 3s, 3p, 3d and 4s atomic states)
and uses 3, 3 and 2 projectors for the s, p and d angular momentum 
channels, respectively.
The O dataset includes 6 valence electrons and 2 projectors in
each of the s and p channels.
The wave functions have been expanded in plane waves up to a 
kinetic energy cut-off of 400 eV for the PtPd system, while a 
higher cut-off of 540 eV has been used for the CoCr systems, 
due to the semi-core states present in the Cr dataset.
Total energy calculations have been performed using the 
Methfessel-Paxton technique with an electronic smearing 
of $\sigma$ = 0.2~eV around the Fermi-level, and using
a self consistency tolerance of $10^{-4}$~eV per simulation 
cell. 
Geometrical optimizations have been done until all force
components on all unconstrained atoms were less than 
$0.01$~eV/\AA.

Spin-polarized (unrestricted) calculations have been performed for all 
CoCr systems and spin-paired calculations for the PtPd systems.
In a few test cases (as reported in Section 3.2.2.), spin unrestricted
calculations for PtPd models all resulted in a spin-paired ground state.
The magnetic ground state of bulk Co in the hcp lattice has been found 
to be ferromagnetic, consistently with the experimental evidence, with a 
computed magnetic moment of 1.65 Bohr Magnetons (to be compared with the 
experimental value of 1.72~$\mu_B$\cite{Kittel}).
For Cr bulk, using a bcc cubic cell we found an antiferromagnetic ground
state and a magnetic moment per atom of 1.02~$\mu_{\textrm{B}}$.
While the experimental magnetic ground state is a longitudinal spin-density 
wave with an incommensurate wave vector~\cite{Corliss_1959},  our magnetic 
solution is in agreement with a number of previous electronic structure 
calculations at the DFT level using a simple bcc cell structure 
(see, e.\ g., Ref.~\cite{Cerny}).

Surfaces were simulated with periodically repeated simulation cells consisting
of a slab of stacked metal layers separated by a vacuum gap of thickness corresponding
to at least six metal layers in the direction perpendicular to the surface.
Distributions of $15\!\times\!15\!\times\!1$, $9\!\times\!9\!\times\!1$ and
$6\!\times\!6\!\times\!1$ k-points have been used for
$1\!\times\!1$, $\sqrt{3}\times\!\sqrt{3}$ and $2\!\times\!2$ surface cells,
respectively.
Oxygen adsorption energies per atom have been calculated either using symmetric
configurations with O atoms adsorbed on both sides of the metal slab, or asymmetric 
configurations with oxygen only on one slab side (for the models in Section 3.2.1).
In the latter case, the influence of the resulting electric dipole on the computed 
total energy values was estimated  to be less than 10~meV per surface atom according 
to standard methods~\cite{Neugebauer}, and has thus been neglected in the presented results. 
The Co-rich CoCr alloy has been modeled by assuming an hcp crystal lattice with
equilibrium lattice constant of pure Co.
In the case of Pt-rich and Pd-rich PtPd alloys, the lattice constants have been set
equal to the equilibrium values obtained for pure Pt and pure Pd, respectively.
Convergence of energy differences with respect to the used cut-off energies,  
k-point grids, surface slab thickness and size of the vacuum gap between slabs 
has been tested in all cases within a tolerance of 10~meV/atom.

Atomic charges have been computed according to a Bader decomposition
of the charge density according to the grid-based algorithm of 
Henkelman et al.~\cite{Henkelman}.
In all cases, the density of the real-space grids used results in errors
on the obtained charge values of the order of 0.01~e.

\section{Alloy surface models}

\subsection{Stacks of single element layers}

We start our investigations focusing on a $1\!\times\!1$ surface cell model
corresponding to a stack of nine (111) atomic planes for PtPd
and of (0001) planes for CoCr, each plane containing only one element.
In the case of PtPd, the considered systems contained either 7 Pt atoms and 2 Pd atoms
(Pt-rich model) or 7 Pd atoms and 2 Pt atoms (Pd-rich model), with the minority element 
placed in symmetric positions with respect to the slab center.
In the case of CoCr, we considered models  with a Co:Cr ratio of 2:1
(typical for technological alloys), where one Cr atom was placed in the center 
of the slab and the other two Cr atoms in symmetric positions with respect to it.
The surface energy per atom of these models has been computed as half the difference 
between the total energy of the system with vacuum gaps separating the 
slab from its repeated images in $z$ direction and the total energy of the 
corresponding system without gap.
The total energy of each system has been computed both for bare and oxygen-covered 
surfaces, relaxing the atomic positions of all atoms composing the surface slab.
The results are summarized in the graphs of Fig.~\ref{1x1_seg}, in which
the different models are labeled on the $x$-axis according to the  
position of the minority element with respect to the surface layer 
(the surface layer is labeled 1, the first subsurface layer is labeled 2,
and so on).

In the presence of an oxygen ad-layer, the calculated total energies are indicative of a 
net driving force for segregation of the less noble atoms Pd and Cr to the surface layer, 
due to their larger affinity to oxygen (Figs.~\ref{1x1_seg}(c,f,i)).
For bare systems, however, the situation is more complex and is addressed
in detail below, starting with the PtPd system.

In the case of the Pd-rich model, the minimum energy is obtained
for a Pd surface layer and the Pt layer in the slab interior.
The surface energy shows a non-monotonous dependence on the Pt layer position,
being highest for a Pt subsurface layer. 
In contrast, for the Pt-rich model the most stable configuration 
corresponds to a Pt surface layer and a Pd subsurface layer, which also
leads to the lowest surface energy.
Given that pure Pd has lower surface energy than pure Pt
(0.56~and 0.64~eV/atom, respectively), at a first glance it may appear 
surprising that the minimum surface energy of the Pt-rich system is obtained for 
a Pt surface layer.
However, this can be explained by looking at the Bader atomic charges 
presented in Table~\ref{t:Bader_charges}. 
In general, the mixing of Pt and Pd in an alloy results in charge transfer
between the atoms.
In the present case, about 0.1 electrons are donated from Pd atoms into Pt atoms.
Therefore, in the Pd-rich system with a Pt subsurface layer, the d-band
of the Pd surface atom becomes slightly depleted of electrons.
Conversely, in the Pt-rich system, the filling of the d-band of the Pt surface 
atom increases due to donation from the Pd subsurface atom (Table~\ref{t:Bader_charges}).
As originally proposed by Friedel~\cite{Friedel} and later discussed
in~\cite{Skriver,Methfessel,Alden}, for metallic systems with more than
half-filled d-bands, the surface energy is expected to decrease with increased 
filling of the d-band.
Our results are in line with this model, clearly indicating that electron-transfer
from the subsurface to the surface layer causes lowering of the surface
energy (subsurface Pd in Pt matrix), whereas electron-transfer in the opposite
direction (subsurface Pt in a Pd matrix) leads to a surface energy increase 
compared to the corresponding bulk values.

When the minority element moves into the inner part of the
slab model, the system is expected to lower its total
energy due to the favorable enthalpy of mixing of
the alloys.
However, the differences in energy in the PtPd systems due to
crystal truncation (Fig.~\ref{1x1_seg}(a,d)) amount to about 20 to 30 
meV per surface atom, i.e.\ to about 50 meV per unit cell.
Since the total energy differences are of the order of 100~meV,
we must conclude that contributions due to crystal truncation
and element mixing compete with equal strength to determine 
the overall stability of the systems.
In particular, the combination of these two effects leads
to the peculiar, well-defined total energy minimum found 
for a Pd subsurface layer in the Pt-rich system.

The same qualitative picture holds for CoCr, where the 
charge transfer effects are more pronounced compared to PtPd.
Cr atoms in a Co matrix donate up to about 0.3 electrons 
into the d-band of Co atoms.
Thus, a minimum surface energy is obtained for a surface Co layer 
and a subsurface Cr layer.
Contrary to PtPd, however, this effect is enhanced by the
lower surface energy of pure Co compared to Cr in an hcp
lattice (by about 0.16~eV per surface atom).
Moreover, the favorable mixing of the two elements acts in 
the same direction as the surface energy. 
Thus, a total energy minimum is obtained for the system 
with Cr in the subsurface layer.
As the Cr layer moves toward the slab center (also consisting
of Cr), the total energy increases, partly because of the 
unfavorable demixing into separated Co and Cr phases.
Additionally, this may also originate from  unfavorable  magnetic
ordering in our $1\!\times\!1$ model, which forces the Cr atoms
within each layer to bear the same spin moment (while in bulk bcc-Cr nearest 
neighbor atoms couple antiferromagnetically with a magnetic moment of 
1.02~$\mu_{B}$).
The situation is particularly unfavorable when the Cr layers are not separated 
by Co layers, to which they are found to couple antiferromagnetically.

\subsection{Mixed layers}

To go beyond the limitations imposed by $1\!\times\!1$ surface models
we consider in this section larger alloy surface models composed of mixed 
atomic layers.
The results obtained in the previous section (Fig.~\ref{1x1_seg}) indicate 
that the differences in surface energy and oxygen affinity are mainly determined
by the elements in the surface and first subsurface layer, while the deeper layers
influence the surface chemistry only to a minor extent.
For this reason, we have investigated surface models taking into account different 
compositions of only the topmost layers.

\subsubsection{CoCr system}

In the case of CoCr, the surface has been modeled by a  $2\!\times\!2$ 
surface cell with a metallic slab of six hcp 
atomic layers separated by a vacuum gap of the same thickness.
Five systems of equal composition (Co:Cr=2:1), but
with different pseudo-random arrangements of Co and Cr atoms on the lattice sites 
have been considered (see Fig.~\ref{f:CoCr_2x2}).
The first system corresponds to surface segregation of Cr only.
Going from the first to the fifth system, the Cr atoms mix
with Co in the inner part of the slab and Co atoms move
to the surface up to pure Co surface segregation.

Electronic structure calculations of these five systems 
reveal transfer of electrons from Cr to Co atoms, similarly as
in the previously discussed $1\!\times\!1$ models
(see Tab.~1).
In the case of bare metal surfaces,  the total energy of 
the systems decreases from system 1 to system 5 by about 3.1~eV 
(Fig.~\ref{f:CoCr_2x2}, top graph).
This is due to a combination of the smaller surface energy
of Co(0001) with respect to hcp-Cr(0001) and of the
negative enthalpy of mixing of the alloy.
For models containing two elements $A$ and $B$, the enthalpy of 
mixing $\Delta H_{\rm{mix}}$ can be computed as:
\begin{equation}
\Delta H_{\rm{mix}} = E_{\rm{AB}}^{\rm{Slab}} - N_{\rm{surf}} \gamma - N_{\rm{A}}E_{\rm{A}}^{\rm{Bulk}}-N_{\rm{B}}E_{\rm{B}}^{\rm{Bulk}},
\end{equation}
where $E_{\rm{AB}}^{\rm{Slab}}$ is the total energy of a given slab model,
$\gamma$ is the corresponding surface energy, $N_{\rm{surf}}$ is the number 
of surface atoms, $E_{\rm{A}}^{\rm{Bulk}}$ and 
$E_{\rm{B}}^{\rm{Bulk}}$ are the bulk energies per atom of the separate elements, 
and $N_{\rm{A}}$ and $N_{\rm{B}}$ are the number of atoms of species A and B in the models.
The surface energy and enthalpy of mixing of the five considered
CoCr models are reported in the graphs of Fig.~\ref{f:se_em_cocr}.
The graphs show a decrease of both surface energy and mixing enthalpy as 
the Cr atoms move in the Co matrix, except for system 5, where 
complete Cr depletion of the surface layer lead to a slight 
increase of $\Delta H_{\rm{mix}}$.
$\gamma$ decreases steadily from 0.93~eV per surface atom 
for system 1 (Cr surface layer) to 0.72~eV per surface atom for system~5 
(Co surface layer).
Consistent with the discussion in the previous section, the surface
energy of system 1 is slightly larger than the surface energy of pure
Cr in the hcp lattice due to electron transfer from surface Cr
atoms to subsurface Co atoms, which depletes the Cr d-band more than 
in the pure metal.

Contrary to the trend observed for the bare models, Cr surface 
segregation becomes energetically favorable in the presence 
of 1~ML of oxygen adsorbed on hollow sites on the alloy surface.
The total energy gain for going from system 5 to system 1 is 1.2~eV per surface 
atom (Fig.~\ref{f:CoCr_2x2}), indicating a strong thermodynamic force for 
oxidation-driven Cr segregation.

As far as the magnetic ordering is concerned, the magnetic moments on 
the Co atoms are about 1.5~$\mu_{B}$ in pure Co layers, and are reduced 
to about 0.6~$\mu_{B}$ in layers containing Cr atoms.
In general, Cr atoms are found to couple antiferromagnetically
to the atoms of the Co matrix. Between Cr--Cr neighbors, antiferromagnetic coupling
is often observed.
The magnetic moments on the Cr atoms vary from 0.0 to 
about $\pm1.0$~$\mu_{B}$, depending on the local alloy structure in a 
non-trivial manner.
In the surface layer, an overall enhancement of the magnetic
moments is found for a bare surface, while the presence of an oxygen
ad-layer leads to almost fully quenching the surface magnetization,
in agreement with previous calculations of similar systems~\cite{Pick}.
For each system, the existence of different magnetic solutions has been checked
by different initialization of the wave functions prior to electronic minimization.
Often the magnetic ordering obtained after electronic minimization changed 
with respect to the initial guess, leading to the same final configuration.
In a few cases different local minima have been found, but the total energy differences 
among different magnetic solutions are about one order of magnitude
smaller than the differences between different atomic configurations (i.e. of the
order of 0.1~eV against a few eV).

\subsubsection{PtPd system}

In the case of the PtPd system, the alloy surface has been modeled by a 
$\sqrt{3}\!\times\!\sqrt{3}$ surface unit cell (containing three surface atoms) 
and a slab of six metal layers separated by an equally thick vacuum gap.
The composition of the first two layers has been varied whereas the 
other four layers, referred to as ``substrate'' in the following,
consisted of either pure Pt or pure Pd.
The Pd:Pt compositions in the first {\it two} layers have been chosen as 2:1, 1:1, and 1:2. 
At each composition, the Pt concentration in the surface layer has been varied to cover 
all possible surface terminations.
Accordingly, the concentration in the subsurface layer has been adjusted to
fix the composition of the two topmost layers.
The total energy of all systems has been calculated for bare surfaces 
and in the presence of an oxygen ad-layer fully relaxing the atomic
positions except for the two bottom layers of the substrate. 
The results of energy calculations for different near-surface compositions 
and substrates are summarized in Fig.~\ref{3x3_seg}.

In the case of a Pd substrate, for all three compositions 
the most stable configuration was obtained for maximum Pd concentration 
in the surface layer.
This is consistent with the results of the Pd-rich $1\!\times\!1$
system (Fig.~\ref{1x1_seg}b).
On the contrary, in the case of a Pt substrate, in all cases the stability of
the system increases with increasing Pt concentration in the surface
layer.
This also is in agreement with the results of the Pt-rich $1\!\times\!1$
model (Fig.~\ref{1x1_seg}e), and reflects both the favorable mixing
of Pt and Pd atoms in the alloy and the lower surface energy of systems
with a Pt surface layer and a Pd subsurface layer.
Confirming the general trend observed for the $1\!\times\!1$ models, the effect 
of an oxygen ad-layer is to stabilize the systems with large Pd concentration 
in the surface layer irrespective of the substrate or the subsurface composition.

\section{Oxygen adsorption}

In this section we discuss in detail the results of the calculated adsorption energies 
of oxygen atoms on different  CoCr and PtPd alloy surfaces.
Within our DFT approach, the adsorption energy (or heat of adsorption) 
is defined as 
\begin{equation}
E_{\rm{ads}} = \frac{1}{N_O} \left[ E_{O@M} -  E_{M} - N_O \frac{1}{2} E_{O_2}  \right]
\end{equation}
where $E_{O@M}$ is the total energy of a given slab model with adsorbed oxygen
and $E_{M}$ the corresponding energy of the bare slab. 
$E_{O_2}$ is the total energy of an oxygen molecule and  
$N_O$ is the number of adsorbed O atoms.
According to our calculations, the fcc hollow sites are the most stable 
adsorption sites for all systems considered (for instance, for the case
of PtPd, adsorption on hcp or bridge sites is less stable by about 0.2 and
0.4--0.5~eV, respectively, at a coverage of 0.33~ML).
On pure Pt(111), Pd(111), Co(0001) and hcp-Cr(0001), the calculated 
adsorption energies at an oxygen coverage of 1~ML amount 
to -0.17, -0.22, \; -1.70, and -3.52~eV, respectively~\cite{O_ads}.
In the case of alloy surfaces, deviations from these values are expected
due to different surface reactivities originating from charge
transfer processes.
In particular, as shown in Table 1, oxygen is more weakly bound to a 
Pt surface layer in the presence of a Pd subsurface layer than to pure Pt.
In the opposite case of a Pt subsurface layer below a Pd surface layer, 
oxygen is more strongly bound than on pure Pd.  
Similarly, for a Cr subsurface layer below a Co surface layer,
oxygen binding is weaker than on pure Co by 0.05~eV, 
whereas for a Co subsurface layer below a Cr surface layer, oxygen binding 
is stronger than on pure Cr by 0.08~eV.

In all considered cases with the same element in the surface layer, 
the trend in the adsorption energy follows the same trend as the
calculated surface energies.
For instance, for the PtPdPd, PtPdPt and PtPtPd layer sequences
(starting from the surface, cf. Table~1),  the adsorption energies are 
-0.05, -0.11, and -0.17~eV, respectively,  and the correspondent 
surface energies are 0.58, 0.59 and 0.63~eV per surface atom.
As a second example,  for the PdPdPt, PdPtPd and PdPtPt layer sequences, 
the adsorption energies are -0.19, -0.22, and -0.27~eV, respectively, and the
correspondent surface energies are 0.56, 0.59, and 0.60~eV.

A similar relationship between oxygen adsorption energy and PtPd 
layer compositions has been found for the mixed layer models at 
an oxygen coverage of 1/3~ML.
The oxygen adsorption energies are presented in Table~\ref{t:adso}
along with the distances between oxygen and metal surface atoms for
different PtPd configurations as well as for the pure elements.
The adsorption energy has been found to be nearly independent of the
underlying Pd or Pt substrate, but to depend substantially on the
surface composition.
As a general trend, the adsorption energy increases with decreasing Pt 
concentration in the surface layer, and is maximum for a fully segregated 
configuration where Pd occupies all three surface sites and Pt all three 
subsurface sites.
Correspondingly, the distance between oxygen and the surface atoms 
decreases with increasing adsorption energy.

The observed behavior is directly related to electron transfer from
less electronegative atoms (Pd or Cr) to more electronegative atoms
(Pt or Co).
These charge transfer effects (see Table~\ref{t:Bader_charges}) result in
shifts of the valence electron band with respect to the Fermi level.
According to the d-band center model introduced by Hammer and 
N{\o}rskov~\cite{Hammer}, increased or decreased band filling leads 
to weaker or stronger oxygen binding, respectively.
This model is found to be reasonably valid for the surface models 
considered here, as shown in Fig.~\ref{d-band-ptpd}(a) for PtPd and  
in Fig.~\ref{d-band-cocr}  for CoCr.
In both cases, the oxygen adsorption energy is roughly linearly 
related to the position  of the center of the density of 
states (DOS) projected on the d-bands of the surface atoms 
of the bare surface models, with respect to the Fermi level.
This highlights the intimate connection between the charge-transfers 
due to alloying and the  resulting surface reactivity.

We found that the different d-orbital components contribute to
a similar extent to the valence band DOS, and that the centers of 
each d-orbital projection vary with alloying in a similar way 
(although the absolute values differ for each component), as 
reported in Fig.~\ref{d-band-ptpd}(b) for the PtPd case.
Moreover, in the case of PtPd the contribution of the metal s- and 
p-states to the O bonding is negligible, and the positions of the
centers of the whole valence bands differ from the centers of the d-band
by about 0.05 eV. 
These differences are larger (about 0.2 eV) in the case of the 
CoCr systems, and are reported in Fig~\ref{d-band-cocr}.
Finally, for CoCr, the surface may exhibit more or less pronounced 
magnetization.
In this case, a roughly linear relation between d-band center and 
oxygen adsorption energy is obtained for the sum of the d-bands
associated with the majority and minority spin-manifolds, but 
not for either separate manifold (Fig~\ref{d-band-cocr}).

\section{Conclusions}

The results presented in the previous sections are indicative of substantial 
variations of the surface energy and of the reactivity towards oxygen adsorption
of both PtPd and CoCr alloys, depending on the composition of 
mainly the first two surface layers.
The alloy properties are generally determined by the electronic interactions
between the components and by strain effects due to different lattice constants.
The latter is, however, unimportant for the present bimetals because of the
very similar lattice constants of the corresponding alloy components. 
The results of the present study suggest that the behavior of the considered
noble and non-noble bimetals can be rationalized by the same electronic 
structure arguments, originating from the electron transfer between the alloy 
elements due to their different electronegativities.
These amount to 1.88 and 1.66 for Co and Cr, and to 2.28 and 2.20 for Pt and Pd, 
according to the Pauling scale~\cite{Webelements}.
Consistently, analysis of the atomic Bader charges in the considered 
alloys revealed a donation of about 0.1\,e from Pd to Pt and of
about 0.3\,e from Cr to Co.
When the more electronegative element is present on the surface at
high concentration and the less electronegative element is present in the 
subsurface layer, then the d-band of the surface atom becomes
more filled compared to the pure element.
This results in a lower surface energy and a weaker oxygen binding  
with respect to the pure metal.
On the contrary, a higher surface energy and stronger oxygen binding is obtained
when the less electronegative atom is in the surface layer and the
more electronegative atom in the subsurface layer.
These findings are in agreement with the d-band center model~\cite{Hammer}
which relates the reactivity of a metal towards oxygen adsorption
to the position of the center of the DOS projected onto the d-orbitals of 
the surface atoms (see Figs.~\ref{d-band-ptpd} and~\ref{d-band-cocr}).

In the case of PtPd alloys, the effect of d-band filling would drive 
segregation of the more electronegative Pt toward the surface.
However, pure Pt presents a higher surface energy than pure Pd,
causing an opposite driving force for Pt segregation toward the bulk.
It is therefore not trivial to predict which element would segregate
at the surface under vacuum conditions.
Indeed, as pointed out in the introduction, existing theoretical
and experimental studies are not conclusive in this respect.
Our results suggest that the surface segregation may depend on the alloy composition,
revealing favorable Pt surface segregation at low Pd concentrations, but a clear preference 
for Pd segregation at high Pd concentration.
This may reconcile experimental findings of Pd surface
segregation~\cite{Moysan,Rousset1,Oetelaar} in PdPt alloys with recent
investigations clearly showing the presence of a ``Pt skin", which
forms on the surface of Pt alloys including less electronegative
elements such as Ni or Co~\cite{Stamenkovic_2007}.
Moreover, our results may help to explain the enhanced efficency for oxygen reduction reactions
of PtPd nanoparticles with a Pd concentration of about 25~\%~\cite{Ye_2007}.

In the presence of adsorbed oxygen, the large affinity of Pd to oxygen
causes preferred Pd segregation in all cases.
Consistently, heating PtPd alloys in oxygen showed a clear enhancement of the 
Pd surface concentration~\cite{Oetelaar}. 
Similarly, in the case of CoCr alloy surfaces exposed to  an oxidizing environment, a 
strong thermodynamical driving force for surface segregation of chromium exists 
because of the large formation enthalpy of chromium oxide compared to cobalt oxides.
Under vacuum conditions, however, both because of the higher electronegativity and 
of the lower surface energy, Co atoms are expected to segregate to the surface of alloys
at the typical technological Cr concentration of ~30~at\,\%.
Given the different favorable surface compositions depending on the environmental
conditions, a question which remains to be answered is how  the rearrangement of 
surface atoms takes place during oxidation of initially bare Co-rich CoCr alloys or 
Pt-rich PtPd alloys.
This issue will be addressed in future work including molecular
dynamics simulations of the oxidation reactions.

\subsection*{Acknowledgments}

The authors would like to thank G. Kresse for his help with the generation of
the PAW datasets used with the VASP code.
Computational resources were provided by the Zentrum f\"ur Informationsdienste 
und Hoch\-leistungsrechnen Dresden, by the Scientific Supercomputing Center
of the University of Karlsruhe, and by the Hochleistungsrechenzentrum Stuttgart. 
NS acknowledges financial support from the Austrian FWF and
is
grateful to the IfWW, TU Dresden, for its hospitality in February and
November 2007. 
LCC acknowledges support by the Deutsche Forschungsgemeinschaft within the Emmy 
Noether Program.
This work has been supported by the Deutsche Forschungsgemeinschaft.

\clearpage

\section*{Tables}

\begin{table}[h!]
\caption{
Bader charges and oxygen adsorption energies E$_{ads}$ for different 
$1\!\times\!1$ layer stacks.
The models are labeled with layer numbers as in the graphs of Fig.~\ref{1x1_seg}.
For clarity, the elements in the first, second and third layer are
also reported (e.\ g. PtPdPd).
The symbols BL$i$ and OL$i$ in the left column denote the 
Bader charge of an atom in the $i$-th layer 
for a bare surface (BL) and in the case of an oxygen ad-layer (OL).
\label{t:Bader_charges}}
{\small
\begin{center}
\begin{tabular*}{\textwidth}{l|ccc|ccc|ccc}
\hline
\hline
    & \multicolumn{3}{c|}{Pd-rich} & \multicolumn{3}{c|}{Pt-rich} & \multicolumn{3}{c}{Co-rich}  \\ 
    &       1              &          2          &        3       
    &       1              &          2          &        3       
    &       1              &          2          &        3       \\
    &  PtPdPd\hskip -1mm\  & PdPtPd\hskip -1mm\  & PdPdPt\hskip -1mm\     
    &  PdPtPt\hskip -1mm\  & PtPdPt\hskip -1mm\  &  PtPtPd\hskip -1mm\   
    &  CrCoCo\hskip -1mm\  & CoCrCo\hskip -1mm\  & CoCoCr\hskip -1mm\   \\
 \hline
BL1  &  -0.09  &   0.02  &  -0.04  &   0.02  &  -0.10   &  -0.04   &   0.15  &  -0.18  &  -0.03  \\
BL2  &   0.09  &  -0.07  &   0.08  &  -0.03  &   0.14   &  -0.02   &  -0.14  &   0.33  &  -0.12  \\
BL3  &  -0.00  &   0.05  &  -0.09  &   0.02  &  -0.05   &   0.11   &  -0.01  &  -0.15  &   0.28  \\
\hline
O    &  -0.72  &  -0.67  &  -0.68  &  -0.68  &  -0.69   &  -0.72   &  -0.75  &  -0.88  &  -0.92  \\
OL1  &   0.69  &   0.80  &   0.74  &   0.81  &   0.69   &   0.78   &   0.86  &   0.76  &   0.93  \\
OL2  &   0.03  &  -0.16  &   0.01  &  -0.11  &   0.12   &  -0.09   &  -0.11  &   0.27  &  -0.17  \\
OL3  &  -0.02  &   0.04  &  -0.12  &  -0.02  &  -0.06   &   0.08   &   0.01  &  -0.14  &   0.30   \\
\hline
\hline   
 E$_{ads}$   & -0.05  & -0.22  &  -0.19 &    -0.27  &  -0.11  &  -0.17 &   -3.60  &  -1.65  & -1.79   \\
\hline \hline
\end{tabular*}
\end{center}
}
\end{table}

\clearpage

\begin{table}[h!]
\caption{ 
Oxygen adsorption energies, $E_{ads}$, at an oxygen coverage of 1/3~ML
as well as oxygen-metal distances on PtPd(111) surfaces for the mixed layer models
on Pd and Pt substrates (values in brackets concern the Pt substrate).  
The Pt fractions in the two topmost layers are given by $x$ 
and $y$, referring to the surface layer (Pt$_{x}$Pd$_{1-x}$) and
the subsurface layer (Pt$_{y}$Pd$_{1-y}$), respectively (cf.\ Fig.~\ref{3x3_seg}).
The values obtained on pure Pd(111) and Pt(111) surfaces are reported 
in the two bottom lines. 
For comparison, values between -1.29 and -1.64~eV have been computed
for oxygen adsorption at a coverage of 0.25~ML on Pt(111)~\cite{Markus,Nicola},
and values of -1.47 and -1.12~eV have been computed in Ref.~\cite{Todorova} 
for O adsorption on Pd(111) at coverages of 0.25 and 0.50~ML, respectively.
\label{t:adso}}
\begin{center}
\begin{tabular}{lccc}
\hline
\hline
 \hfill\ $x$,$y$ \hfill\  &  $E_\mathrm{ads}$ (eV) &   $d_\mathrm{Pd-O}$ (\AA) &  $d_\mathrm{Pt-O}$ (\AA) \\
 \hline
  0.00, 0.67  & $-1.24$ ($-1.25$) & 2.01 &       \\
  0.33, 0.33  & $-1.19$ ($-1.20$) & 2.06 & 1.97  \\
  0.67, 0.00  & $-1.10$ ($-1.11$) & 2.09 & 2.00  \\
 \hline
  0.00, 1.00  & $-1.26$ ($-1.26$) & 2.01 &       \\
  0.33, 0.67  & $-1.18$ ($-1.19$) & 2.05 & 1.97  \\
  0.67, 0.33  & $-1.08$ ($-1.08$) & 2.09 & 2.01  \\
  1.00, 0.00  & $-0.99$ ($-1.00$) &      & 2.04  \\  
\hline
  0.33, 1.00  & $-1.23$ ($-1.24$) & 2.05 & 1.97  \\ 
  0.67, 0.67  & $-1.10$ ($-1.10$) & 2.09 & 2.01  \\
  1.00, 0.33  & $-0.94$ ($-0.95$) &      & 2.05  \\
 \hline
  0.00, 0.00  & $-1.21$ \hfill\          & 2.01 &       \\
  1.00, 1.00  &       \hfill\ ($-1.08$)  &      & 2.05  \\
\hline \hline
\end{tabular}
\end{center}

\end{table}%

\vfill\

\clearpage
\section*{Figure Captions}

\begin{figure}[h!]
\caption{\label{1x1_seg}
Calculated surface energies $\gamma$ per surface atom (top row)
and total energies $E$ of PtPd and CoCr layer stacks for bare surfaces
(middle row) and in the presence of an oxygen ad-layer (bottom row).
For each system, the energy $E$ is reported relative to the energy calculated 
for the minority element placed in the fourth layer below the surface.
The layer numbers on the $x$-axis are referred to the surface
layer (labeled 1), so that the first subsurface layer is labeled 2, and 
so on.}
\end{figure}

\begin{figure}[h!]
\caption{\label{f:CoCr_2x2}
Total energies $E$ of different CoCr alloy models (Co bright, Cr dark) 
for a $2\times 2$ (0001) surface unit cell calculated in the absence 
and in the presence of an oxygen ad-layer at a coverage of 1~ML.
The energy values are relative to system 5.}
\end{figure}

\begin{figure}[h!]
\caption{\label{f:se_em_cocr}
Calculated enthalpies of mixing $\Delta H_{\rm{mix}}$, relative to that of
system 1, and surface energies~$\gamma$ per surface atom of the
$2\!\times\!2$ CoCr systems displayed in Figure ~\ref{f:CoCr_2x2}.}
\end{figure}

\begin{figure}[h!]
\caption{\label{3x3_seg}
Calculated total energies $E$ of PtPd(111) surface models, using
a $\sqrt{3} \times \sqrt{3}$ unit cell including 3 surface atoms, 
in the absence (top row) and in the presence (bottom row) 
of 1~ML adsorbed oxygen.
The Pt:Pd composition in the two topmost surface layers changes from left to right
as 1:2, 1:1 and 2:1. 
For each composition, the Pt concentration in the surface layer is varied. 
The four underlying substrate layers consist either of pure Pt (dashed lines)
or pure Pd (full lines).
For each system, the energy $E$ is referred to the total energy obtained
for the maximum Pt concentration in the surface layer.}
\end{figure}

\newpage

\begin{figure}[t]
\caption{\label{d-band-ptpd}
(a) Calculated oxygen adsorption energies versus average positions of the
d-band center projected on the surface atoms of PtPd alloy models.
The points are labeled with $\,\!$ ($x$, $y$) according to the fraction
of Pt in the surface ($x$) and subsurface ($y$) layer (cf.\ Table~\ref{t:adso}).
The dashed line is a linear best-fit to the reported data.
The reported d-band center values refer to bare metal surfaces;
additional shifts of the order of -0.4~eV occur upon adsorption 
of oxygen.
(b)  Calculated oxygen adsorption energies versus average positions of 
the centers of the different d-band components projected on the surface 
atoms of selected PtPd alloy models (labeled as in (a)).
The insert displays the DOS relative to the considered d-band components 
for the (0.00, 1.00) system.
}
\end{figure}

\begin{figure}[h!]
\caption{\label{d-band-cocr}
Calculated oxygen adsorption energies versus centers of the
d-band (filled circles) and of the whole valence band (empty
circles) projected on the surface atoms for the five CoCr alloy 
models shown in Fig.~\ref{f:CoCr_2x2}.
The d-band centers are computed for the DOS of the separate minority and 
majority spin manifolds (upright and downright triangles connected with 
dashed lines) as well as for their sum (full line).}
\end{figure}

\vfill\

\clearpage


\begin{center}
\includegraphics[width=\textwidth]{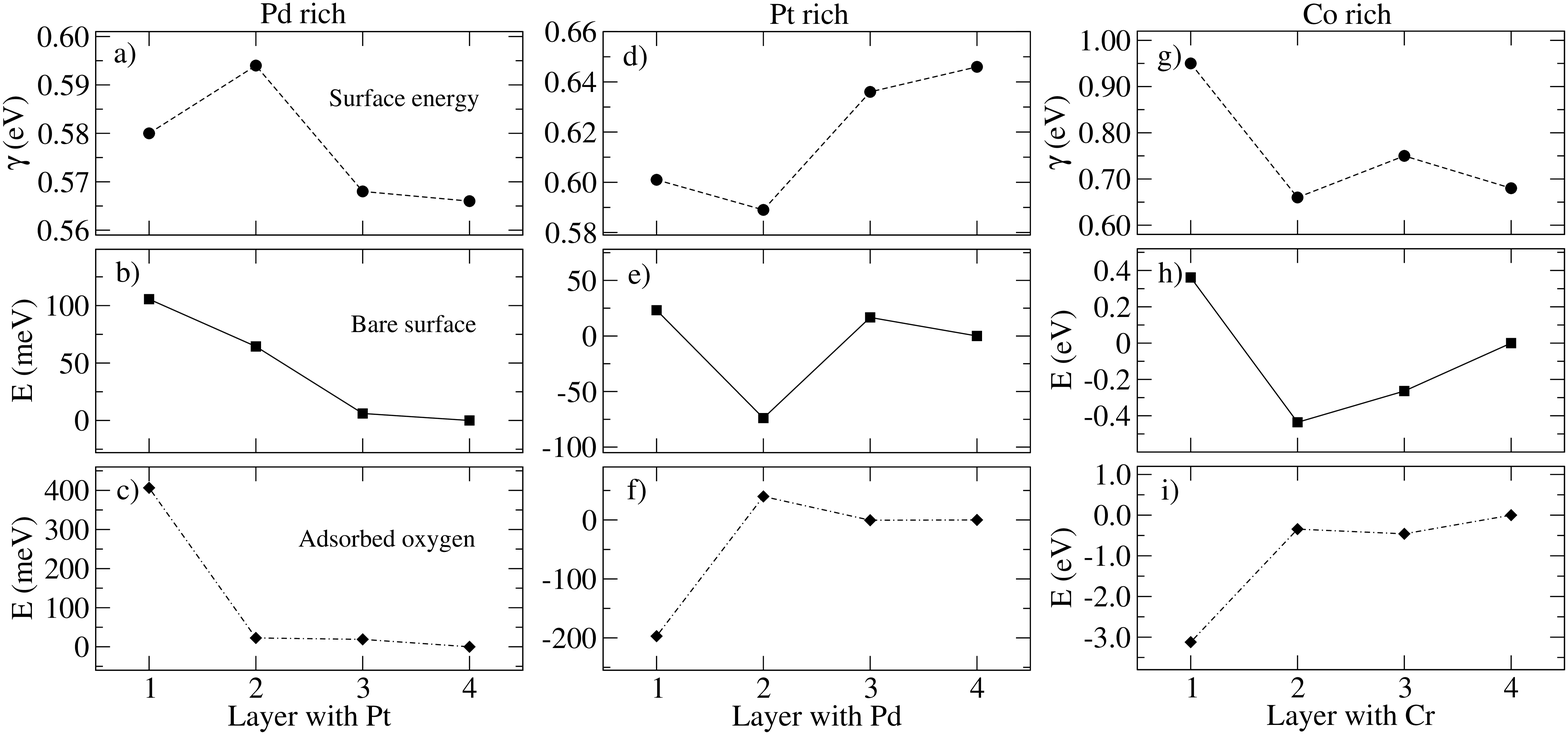}
\vfill
A. Dianat et al., Fig.~\ref{1x1_seg}
\end{center}

\clearpage

\begin{center}
\includegraphics[width=0.45\textwidth]{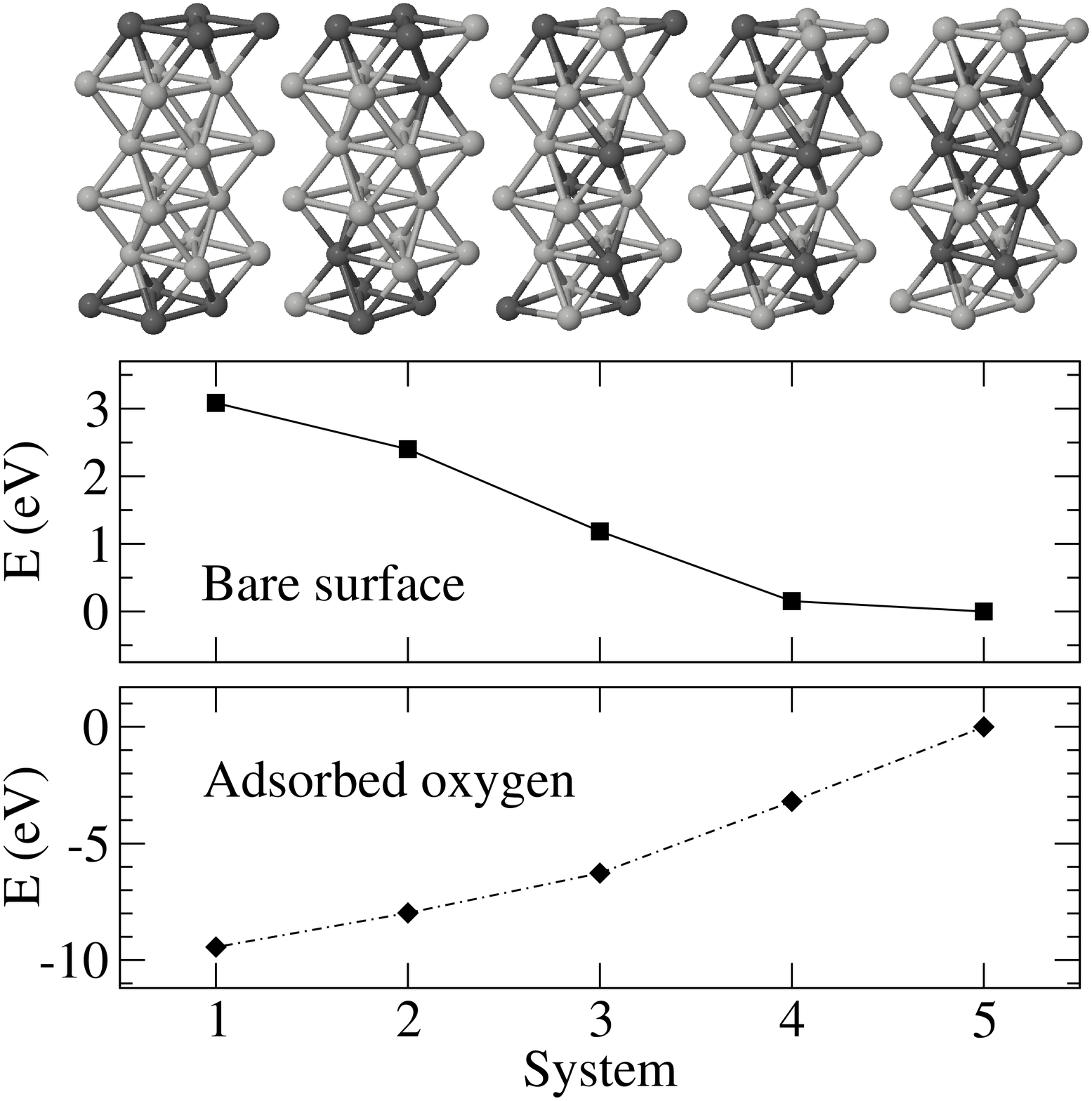}
\vfill
A. Dianat et al., Fig.~\ref{f:CoCr_2x2}
\end{center}

\clearpage

\begin{center}
\includegraphics[width=0.45\textwidth]{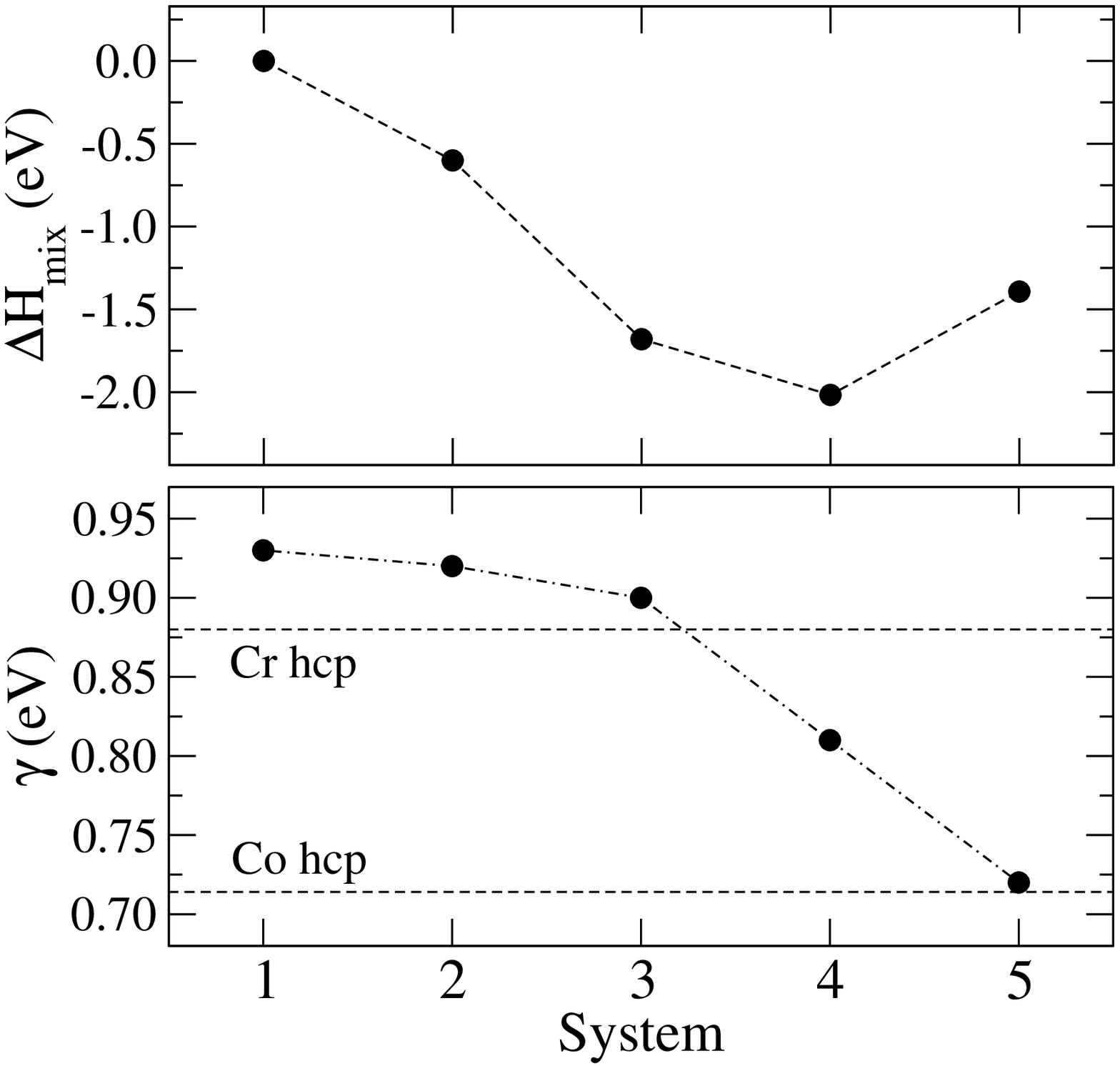}
\vfill
A. Dianat et al., Fig.~\ref{f:se_em_cocr}
\end{center}

\clearpage

\begin{center}
\includegraphics[width=\textwidth]{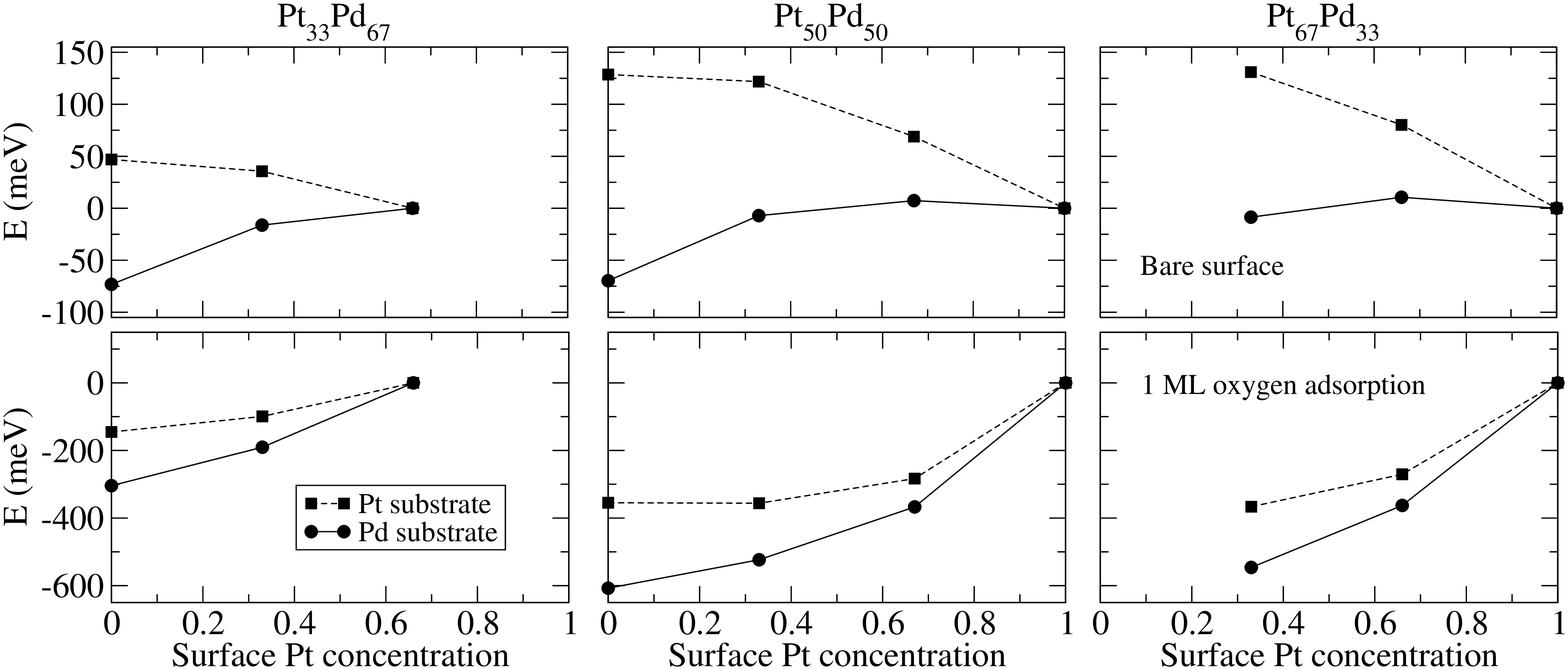}
\vfill
A. Dianat et al., Fig.~\ref{3x3_seg}
\end{center}

\clearpage

\begin{center}
\includegraphics[width=\textwidth]{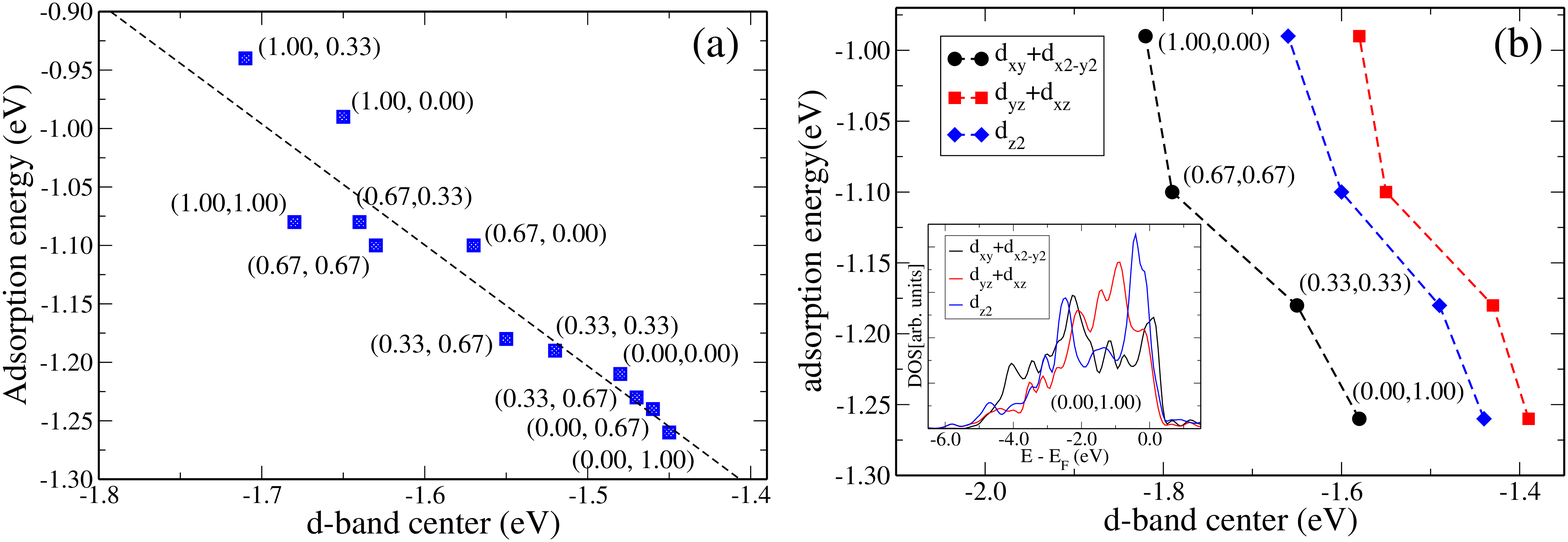} \hfill
\vfill
A. Dianat et al., Fig.~\ref{d-band-ptpd}
\end{center}

\clearpage

\begin{center}
\includegraphics[width=0.46\textwidth]{Fig6.eps}
\vfill
A. Dianat et al., Fig.~\ref{d-band-cocr}
\end{center}


\begin{thebibliography}{99}
\bibitem{O_ads}
Our values are slightly smaller than some of the existing values 
obtained within DFT for Pt(111)  (-0.59~\cite{Nicola}) or Pd(111) 
(-0.38~eV~\cite{Todorova}). These differences are expected to
originate from the large sensitivity of the total energy of the
O$_2$ molecule (which is used here as a reference) to the
actual computational details.
\end{thebibliography}
\end{document}